\documentclass[twocolumn]{openjournal}

\usepackage{graphicx}
\usepackage{latexsym,amssymb}
\usepackage{amsmath,morefloats}
\usepackage[backref,breaklinks,colorlinks,citecolor=blue]{hyperref}
\usepackage{natbib,graphicx,amsmath,color,xcolor}
\usepackage{verbatim}
\usepackage{threeparttable}
\usepackage{xspace}
\usepackage{lmodern}
\usepackage{slantsc}

\newcommand{\mdet}{\textsc{metadetection}\@\xspace}
\newcommand{\mcal}{\textsc{metacalibration}\@\xspace}

\newcommand{\dmcal}{\textsc{deep-field~metacalibration}\@\xspace}
\newcommand{\dmdet}{\textsc{deep-field~metadetection}\@\xspace}
\newcommand{\Dmcal}{\textsc{Deep-field~metacalibration}\@\xspace}
\newcommand{\DMcal}{\textsc{Deep-field~Metacalibration}\@\xspace}
\newcommand{\noshear}{\texttt{noshear}\@\xspace}
\newcommand{\onep}{\texttt{1p}\@\xspace}
\newcommand{\onem}{\texttt{1m}\@\xspace}
\newcommand{\twop}{\texttt{2p}\@\xspace}
\newcommand{\twom}{\texttt{2m}\@\xspace}
\newcommand{\galsim}{\textsc{galsim}\@\xspace}
\newcommand{\descwl}{\textsc{WeakLensingDeblending}\@\xspace}
\newcommand{\cosmodctwo}{\textsc{CosmoDC2}\@\xspace}
\newcommand{\buzzard}{\textsc{Buzzard}\@\xspace}

\xspaceaddexceptions{+}

\shorttitle{Deep-field Metacalibration}
\shortauthors{Zhang, Becker, \& Sheldon}

\begin{document}
\title{Deep-field Metacalibration}

\author{Zhuoqi (Jackie) Zhang}
\affil{Department of Astronomy and Astrophysics, University of Chicago, Chicago, IL 60637, USA}
\author{Matthew R. Becker}
\affil{High Energy Physics Division, Argonne National Laboratory, Lemont, IL 60439, USA}
\author{Erin S. Sheldon}
\affil{Brookhaven National Laboratory, Bldg 510, Upton, New York 11973, USA}

\begin{abstract}

    We introduce \textit{\dmcal}, a new technique
    that reduces the pixel noise in \mcal estimators of weak lensing shear
    signals by using a deeper imaging survey for
    calibration. In standard \mcal, a small artificial shear is applied to the
    observed images of galaxies in order to estimate the response the
    object's shape measurement to shear, which is used to calibrate statistical
    shear estimates.  As part of a correction for the effect of shearing correlated noise
    in the image, extra noise
    is added that increases the uncertainty on statistical shear
    estimates by $\sim 20$\%.  Our new \dmcal technique leverages a
    separate, deeper imaging survey to calculate calibrations with less
    degradation in image noise.  We demonstrate that weak lensing shear
    measurement with \dmcal is unbiased up to second-order shear
    effects for isolated sources. For the Vera C. Rubin Observatory
    Legacy Survey of Space and Time (LSST), the improvement in weak lensing
    precision will depend on the somewhat unknown details of the LSST Deep
    Drilling Field (DDF) observations in terms of area and depth, the relative
    point-spread function properties of the DDF and main LSST surveys, and the
    relative contribution of pixel noise versus intrinsic shape noise to the
    total shape noise in the survey. We conservatively estimate that the
    degradation in precision is reduced from 20\% for \mcal to $\lesssim5$\% for
    \dmcal, which we attribute primarily to the increased source
    density and reduced pixel noise contributions to the overall shape noise.
    Finally, we show that the technique is robust to sample variance in the
    LSST DDFs due to their large area, with the equivalent calibration
    error being $\lesssim0.1\%$. The \dmcal technique provides higher
    signal-to-noise weak lensing measurements while still meeting the stringent
    systematic error requirements of future surveys for isolated sources.

\end{abstract}

\section{Introduction}\label{sec:intro}

Modern surveys that aim to measure weak lensing signals will constrain a wealth of
fundamental physics, but in order to do so they must meet stringent requirements on the control of systematic
effects (see, e.g., \citeauthor{MandelbaumReview}~\citeyear{MandelbaumReview} for a
review). One of the most daunting issues has been accurately measuring weak
gravitational lensing shears from survey data. For the Vera C. Rubin Observatory Legacy
Survey of Space and Time (LSST), these signals must be measured with systematic
errors no worse than about one part per thousand to avoid degrading cosmological
constraints \citep{huterer2006,descsrd}. A large number of systematic effects and
measurement biases must be overcome to reach this goal, including correcting for the
point-spread function (PSF), noise biases, model biases, selection effects, blending
effects, and detection effects (see, e.g.,
\citeauthor{MandelbaumReview}~\citeyear{MandelbaumReview} for a comprehensive discussion
and references). So far, the community has developed three techniques
that can reach the part-per-thousand accuracy required by LSST without direct
calibration from image simulations. These techniques are BFD \citep{BernBFD2016,ba14},
FPFS \citep{li2018fpfs,li2022fpfs}, and \mcal \citep{SheldonMcal2017,HuffMcal2017},
all of which meet LSST requirements for unblended, isolated sources. The newly
introduced \mdet technique promises to reach part-per-thousand accuracy even in the
presence of object detection and blending \citep{SheldonMdet2022}. The FPFS method from
\citet{li2022fpfsblend} is almost as accurate in the presence of blending as well. However,
much work remains to fully realize the potential of these techniques in realistic survey scenarios,
including redshift-dependent shear, blending, and detection. These effects may require
simulation-based calibrations \citep[see, e.g.,][]{desy3imsims}.

Despite their success, the \mcal and \mdet techniques as currently implemented
have a nontrivial drawback. In order to reach the required accuracy in shear
measurement, they effectively double the pixel noise variance in the survey images
($\sqrt{2}$ in the noise standard deviation)
or equivalently reduce the signal-to-noise of every source by
about 30\%. As described below, this
effect comes from a numerical correction needed to account for sheared pixel
noise in the \mcal pipeline. This feature results in a loss of precision in the
statistical shear measurements at the $\sim 20\%$ level \citep{SheldonMcal2017}.

Interestingly, modern surveys typically have a wide-field imaging campaign, which forms
the survey used for weak lensing shear measurements, and a deep-field imaging campaign,
which is a much smaller region that is surveyed longer, eventually reaching exposure times
of $10\times$ or more than the main wide-field survey.\footnote{We assume both the deep- and wide-field surveys are calibrated to the same flux scale.} See for example the Dark Energy
Survey deep-fields work \citep{DESDeepFields} or the planned LSST Deep Drilling Fields
(DDFs) \citep{lsst-ddf-design,lsst-ddf-depth,ivezic2019lsst}. Although \mcal currently
does not take advantage of deep-field data, the BFD estimator explicitly uses it,
avoiding increases in the wide-field image pixel noise in the analysis.

In this work, we propose a new technique, \textit{\dmcal},
which greatly reduces noise degradation by taking advantage of a
deep-field survey to derive calibrations. The standard \mcal estimator for
a mean shear has two
components, the uncalibrated shape measurements $\langle \mathbf{g} \rangle$ and the
mean \mcal response matrix $\langle\mathbf{R}\rangle$, both of which are computed from
the wide-field survey data. For a mean shear, these quantities are combined to
form the final shear estimate as
\citep{HuffMcal2017,SheldonMcal2017}
\begin{equation*}
\hat\gamma \equiv \langle\mathbf{R}\rangle^{-1}\langle \mathbf{g} \rangle\ .
\end{equation*}
The \dmcal estimator works by computing the mean \mcal response matrix
$\langle\mathbf{R}\rangle$ from the deep-field survey data instead of the wide-field
survey data. We show below that with this change, we can carefully arrange the
\dmcal estimator to double the effective pixel noise variance in the
deep-field survey data only. Since the deep-field survey data is lower-noise than the
wide-field survey data in the first place, the final \dmcal estimator achieves
higher precision than the original \mcal technique. We find that in all cases this
estimator is as accurate as the original \mcal estimator, achieving better than
part-per-thousand accuracy in the overall shear measurement for isolated sources. We demonstrate that with
this new technique, we will decrease the pixel noise in \mcal images by $\approx30\%$,
increase the signal-to-noise of \mcal weak lensing sources and decreasing their shape noise.
The exact change in the pixel noise depends in detail on the relative exposure time and PSF
distributions of the wide- and deep-field surveys. We further estimate an at least
$\approx15\%$ increase in the statistical precision of weak lensing analyses with
Rubin LSST due to the decreased pixel noise. We conservatively estimate that for
\dmcal the uncertainty in the shear estimator is degraded by $\lesssim5\%$
rather than the approximately 20\% degradation in standard \mcal.

The primary concern when utilizing only the deep-field data to compute the \mcal
response is sample variance. The deep-field survey typically covers a much smaller
area of the sky than the wide-field survey. The \mcal response depends on the properties
of the galaxies from which it is computed (e.g., their profiles, sizes, etc.) and so
will inherit sample variance in those properties due to large-scale structure. We
estimate this effect with two complementary mock catalogs, finding that for the large
$\approx35$ deg$^2$ area of the LSST DDFs \citep{lsst-ddf-design,ivezic2019lsst}, the
resulting sample variance scatter in the \mcal response will be in the range of 0.05\%
to 0.1\%. This level meets LSST requirements, and extensions of our technique that
reweight the deep-field galaxy populations to match the wide-field ones may reduce this
sample variance further.

Our paper is organized as follows. In Section~\ref{sec:math}, we review standard \mcal
and introduce the formalism for \dmcal. Then in Section~\ref{sec:results}, we
cover the main results of our work, including tests of shear calibration (\ref{sec:mc})
and estimates of the gains in signal-to-noise (\ref{sec:sizedepth}). In
Section~\ref{sec:terms}, we present a discussion of the relative importance of various
corrections in the \dmcal estimator. We address the issue of sample variance
in the deep-fields in Section~\ref{sec:sv}. Finally, in Section~\ref{sec:conc}, we
summarize our results and suggest avenues for further research.

\section{\DMcal Formalism}\label{sec:math}

In this section, we introduce the underlying algorithms of \dmcal and describe
one method to apply those algorithms to a realistic survey composed of wide- and
deep-field imaging datasets. We start by examining standard \mcal images in our
formalism. We then use the intuition developed there to motivate correction terms that
can be applied to the deep-field and wide-field images to enable shear measurement.
Finally, we specify the details of how we combine the deep- and wide-field image sets in
practice, including how to make selections on the sources and compute the deep-field
\mcal response.

For conciseness, we adopt the following notation. We denote a convolution
operation of two images as $I_1 \star I_2$, deconvolution as $I_1/I_2$,
and pointwise addition or subtraction as $I_1\pm I_2$. We also define two special
operators on images. First, we define a shearing operator $Y[I, \gamma]$ that shears
the image $I$ by the shear $\gamma$. Second, we define a \mcal operator,
$M[I, N, P, P_r]$, defined as
\begin{eqnarray} \label{eq:mcalop}
\lefteqn{M[I, N, P, P_r, \gamma]=} & & \\\nonumber
& &\ \ Y[I/P, \gamma] \star P_r + Y[N/P, -\gamma] \star P_r
\end{eqnarray}
where $I$ is an image, $N$ is a second image (usually pure noise), $P$ is the image PSF,
$P_r$ is the \mcal reconvolution PSF, and $\gamma$ is a shear. This operator performs
the standard steps of the \mcal algorithm, including the correction for sheared noise as
described in \citet{SheldonMcal2017}. We have not specified the exact
numerical implementation of these operators for pixelated images. In practice, we rely
on the \galsim package which provides high-quality, performant implementations
\citep{GALSIM2015}. Depending on the context, these implementations rely on fast-Fourier
transforms and interpolations of various kinds and orders. Our implementation is
publicly available.\footnote{\url{https://github.com/beckermr/deep-metacal}}

In what follows, we work with single images for both the wide and deep surveys.
Modern surveys use a strategy of covering an area of the sky with overlapping images taken
at different epochs.  Multi-epoch data can be handled by forming postage stamp
coadds around each object \citep{psc} or for \mdet, by using coadds without PSF
discontinuities \citep{BeckerMdet2022,SheldonMdet2022}. These data products
have well defined PSFs, and the coadd noise properties can be determined by
running pure noise images through the coaddition process. The algorithms below
generalize to these multi-epoch scenarios.

\subsection{Standard \mcal}

The standard \mcal estimator is based on a linear expansion of the shape of an object,
$\mathbf{e}$, in the true $\gamma$,
\begin{eqnarray}
\mathbf{e} & = & \left. \mathbf{e} \right|_{\gamma=0}
+ \left.\frac{\partial\mathbf{e}}{\partial\gamma}\right|_{\gamma=0} \gamma
+ {\cal O}(\gamma^2)\nonumber\\
& \equiv & \left. \mathbf{e} \right|_{\gamma=0}
+ \mathbf{R} \gamma
+ {\cal O}(\gamma^2)\nonumber
\end{eqnarray}
where we have defined the response matrix $\mathbf{R}$ as
\begin{equation}
R_{ij} \equiv \left.\frac{\partial e_i}{\partial\gamma_j}\right|_{\gamma_j=0}\ .
\end{equation}
Under the assumption that the intrinsic shapes of objects average to zero,
$\langle \left. \mathbf{e} \right|_{\gamma=0} \rangle = 0$, then we get to linear order
\begin{equation*}
\langle\mathbf{e}\rangle \approx \langle \mathbf{R} \gamma\rangle,
\end{equation*}
which demonstrates that true shear is weighted by the response
in the mean ellipticity.  We can then calculate the response weighted mean shear
\begin{equation} \label{eq:meanshear}
    \hat{\gamma} \equiv \langle \mathbf{R} \rangle^{-1} \langle \mathbf{e} \rangle \approx \langle \mathbf{R} \rangle^{-1} \langle \mathbf{R \gamma} \rangle.
\end{equation}

In standard \mcal the mean response $\langle \mathbf{R} \rangle$ is computed by applying
small artificial shears to the observed images of objects and using a finite difference
derivative to measure how their shapes respond to the change. Including the shape measurements where no
shear has been applied, this process generates five shape measurements per object. These
are the one with no shear, which we denote as \texttt{noshear},
$\pm\epsilon_1$ which we denote \texttt{1p} and \texttt{1m}, and
$\pm\epsilon_2$ which we denote \texttt{2p} and \texttt{2m}.
A numerical response per object can be constructed via
\begin{equation} \label{eq:standard_R}
R_{ij} \approx \frac{e_i^{+j} - e_i^{-j}}{2\epsilon}\ .
\end{equation}
where we have denoted the shape measurement as $e_i^{\pm j}$ for the $i$th shear
component with  with applied shear $\pm\epsilon_j$ on the $j$th axis. The responses per
object are typically quite noisy and so the recommended estimator for the mean shear of
a region of the sky averages the response and shapes separately as above so that
\begin{equation} \label{eq:avgR}
\langle R_{ij} \rangle = \frac{\langle e_i^{+j} \rangle - \langle e_i^{-j} \rangle}{2\epsilon}\ .
\end{equation}
The averages $\langle\ \rangle$ above are taken over each of the five shapes above
separately after all selections have been applied in order to correct for selection
effects. That is, we aggregate the shape measurements from all of the \noshear images,
all of the \texttt{1p} images, etc. into five separate catalogs, apply the object
selections to those catalogs separately, compute the average shape over those catalogs
separately, and then finally compute the mean shear and response from those averaged
shears. This procedure is the one described in \citet{SheldonMdet2020} for \mdet but
applies equally well to \mcal. However, for standard \mcal, when using a fixed
catalog, one can use separate shear and selection
responses with a single set of selections as done in \citet{SheldonMcal2017}. When
applied to isolated objects, this technique can achieve the part-per-thousand accuracy
required by LSST \citep{HuffMcal2017,SheldonMcal2017}.

\begin{figure*}
  \centering
  \includegraphics[width=\linewidth]{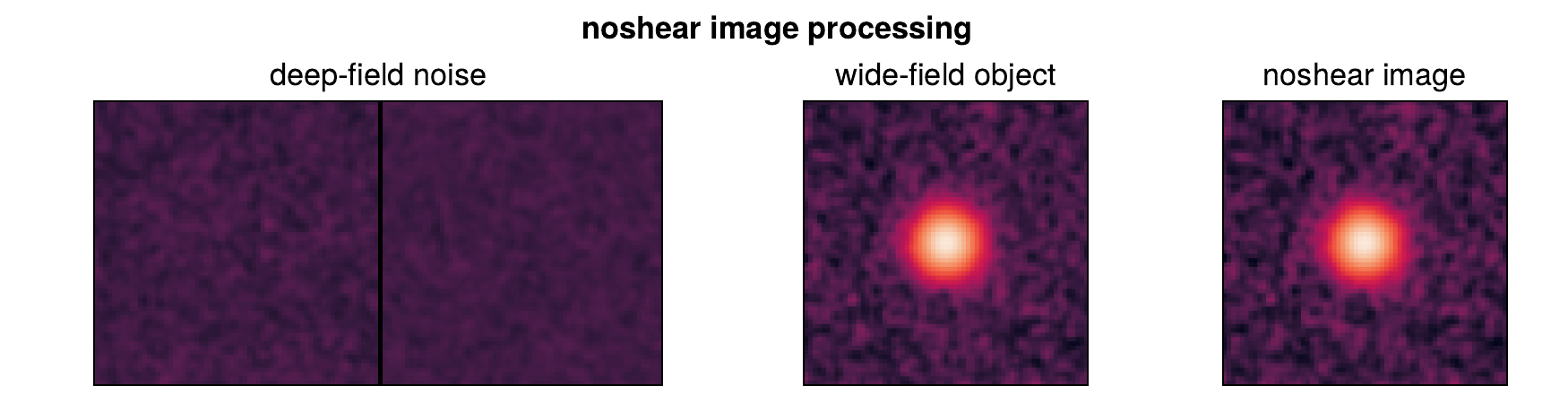}
  \includegraphics[width=\linewidth]{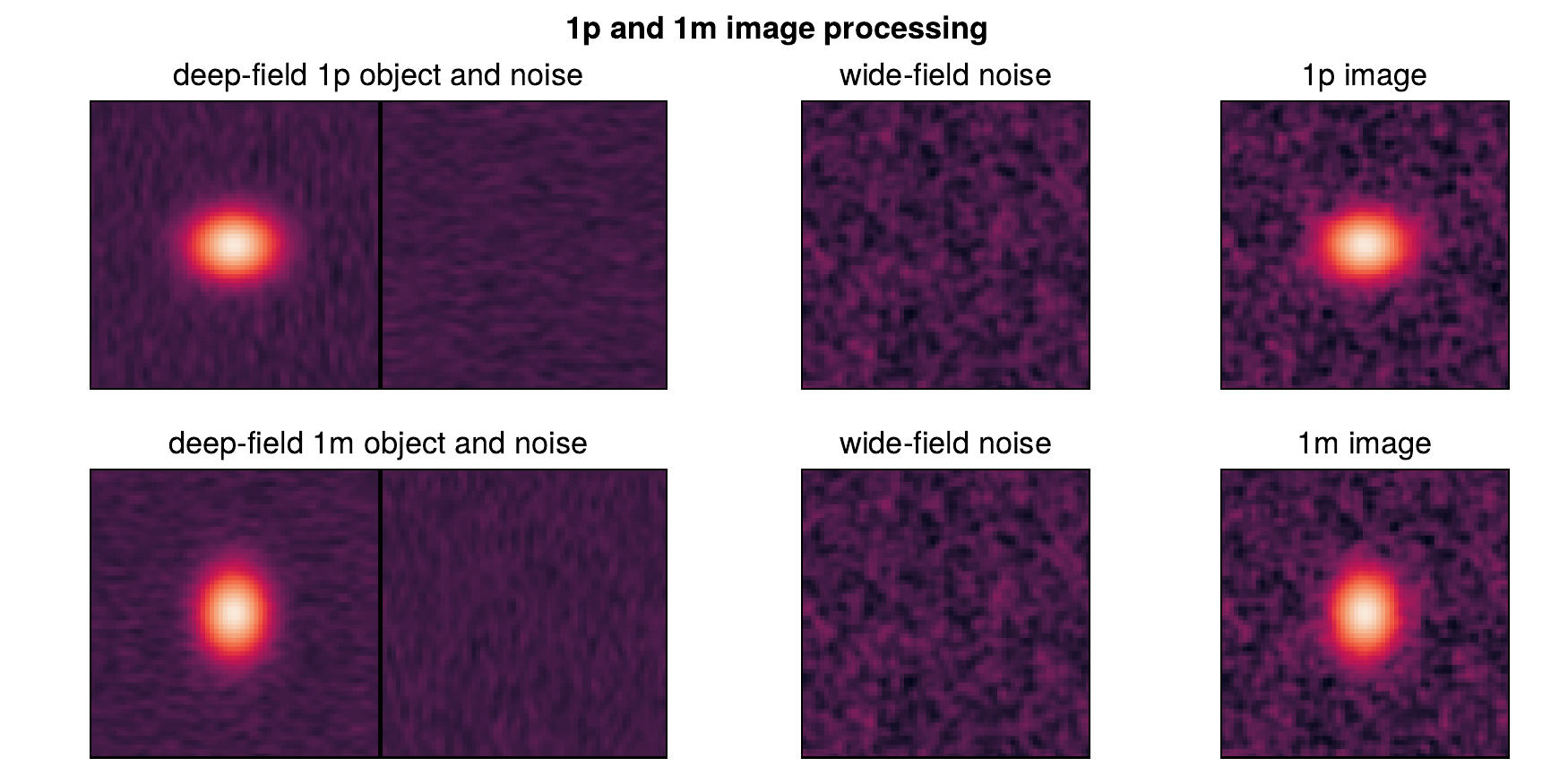}
  \caption{
    \DMcal Image Processing for the \noshear, \onep, and \onem Cases. The top, middle, and bottom rows show the
    \noshear, \onep, and \onem cases respectively. In the \onep and \onem cases, we have exaggerated the artificial
    shear applied to the image. The left column shows the pair of images with the deep-field noise level, either
    $C_{w}$ for the \noshear case or $M[I_{d}, N_{d}', P_{d}, P_{r}, \epsilon]$ for the \onep and \onem cases. The
    middle and bottom pair of images in the left column clearly show the anisotropically correlated noise induced by
    shearing the original image and the oppositely correlated noise in the correction image.
    The middle column shows the image with the wide-field noise level, either $I_{w}/P_{w} \star P_{r}$ for the
    \noshear case or $C_{d}$ for the \onep and \onem cases. The right column for each row shows the total image
    $\hat I_{w,d}$ created by summing the images in left and middle columns. These are the images used by
    \dmcal for computing the uncalibrated shear signal and its response.
  } \label{fig:dmcal}
\end{figure*}

The numerical implementation of standard \mcal has two key features which are relevant
to the discussion of \dmcal below. First, in order to artificially shear an image, one must first deconvolve
the PSF. This procedure is most easily done in Fourier space where it is a simple
division. However, numerical instabilities can arise since one ends up dividing by small
numbers at Fourier modes where the original PSF is nearly zero. To rectify this, one
typically reconvolves the image with a slightly larger PSF, called the
\textit{reconvolution PSF}, which controls these numerical instabilities by suppressing
the affected Fourier modes.

Second, as discussed in \citet{SheldonMcal2017}, one must correct for the effects of sheared background noise
when estimating the response to shear via finite differencing. When applying an
artificial shear to a noisy image, the image is deconvolved and sheared,
which produces an image with an anisotropic noise power spectrum.
This effect, if left uncorrected, biases the shape measurements
for the response images \texttt{1p}, \texttt{1m}, etc. relative to those from the
\noshear images, producing a biased shear estimator (see Equation \ref{eq:meanshear}).
\citet{SheldonMcal2017} tried a
variety of schemes to correct for this effect, but settled on the following procedure.
Before shape measurement under an artificial shear $\epsilon_j$, a pure noise image with
the same noise spectrum as the original image is rotated by 90 degrees, put through the same
deconvolution plus shearing plus reconvolution process, rotated back, and then added
to the sheared original image. This procedure
has the net effect of applying a shear of $-\epsilon_j$ to the noise image, but
works even for distorted images.\footnote{This procedure was developed as part of the
\galsim package \citep{GALSIM2015}.} As detailed below, this procedure cancels the
biases in the shape measurements but has the net effect of doubling the
pixel noise variance in the image. It is this inefficiency we are seeking to rectify in
this work.

Before describing \dmcal, it is useful to cast these two numerical operations into the
notation introduced above. For this, let's denote the galaxy as $G$, the
PSF as $P$, our noise field as $N$, and the \mcal reconvolution PSF as $P_r$. In this
notation, the observed image of an object is $I = G\star P + N$. That is, the image
of the galaxy $G$ is convolved by a PSF $P$ giving $G\star P$, and has additional noise
which results in $G\star P + N$. For a shear $\epsilon$ applied the 1-axis (2-axis), we
have for the \noshear, \onep (\twop), and \onem (\twom) images
\begin{eqnarray}
  \lefteqn{M[I, N', P, P_r, \phantom{-}0]} & & \nonumber \\
    & &=Y[G, \phantom{-}0] \star P_r + Y[N/P, \phantom{-}0] \star P_r + Y[N'/P, \phantom{-}0] \star P_r \nonumber \\
    \lefteqn{M[I, N', P, P_r, +\epsilon]} & & \nonumber \\
      & &=Y[G, +\epsilon] \star P_r + Y[N/P, +\epsilon] \star P_r + Y[N'/P, -\epsilon] \star P_r \nonumber \\
  \lefteqn{M[I, N', P, P_r, -\epsilon]} & & \nonumber \\
    & &=Y[G, -\epsilon] \star P_r + Y[N/P, -\epsilon] \star P_r + Y[N'/P, +\epsilon] \star P_r \nonumber\ .
\end{eqnarray}
The first two terms of the right side of each equation arise from the deconvolution,
shearing, reconvolution process applied to the original image of the object
($(G+N) \star P$). The last term on the right side is the sheared noise correction.
The more formal approach of the equations above illustrates an import numerical
relationship between the noise distributions in the \noshear versus \onep and \onem
(or \twop and \twom) images. Namely, the \noshear image has noise that looks like
\begin{displaymath}
  N_{\rm noshear} = Y[N/P, \phantom{-}0] \star P_r + Y[N'/P, \phantom{-}0] \star P_r
\end{displaymath}
whereas the \onep and \onem (or \twop and \twom) images have identical noise that looks
like
\begin{displaymath}
  N_{\rm 1p,2p} = Y[N/P, +\epsilon] \star P_r + Y[N'/P, -\epsilon] \star P_r\ .
\end{displaymath}
This set of relationships is key to canceling biases in the response do the sheared noise
while also properly accounting for selection effects \citep{SheldonMcal2017}.

For \dmcal, we will need to preserve these relationships while additionally accounting for
the differing PSFs and noise distributions in the wide- versus deep-field images. If
we did not, then selections made on shear sources made in the deep-fields would not match
the wide-fields or vice versa. Similarly, the noise biases would differ between the wide-
and deep-fields, and thus ruin any response-like corrections we would attempt to make. We
demonstrate how to achieve these goals next.

\subsection{\DMcal}\label{sec:deepmcal}

Before presenting the \dmcal technique in detail, it is useful to discuss general constraints
on the method and to develop some intuition. First, as discussed above, we are seeking to avoid
doubling the wide-field image variance. This doubling comes from corrections for numerical operations
that shear pixel noise. Therefore, our final technique should avoid shearing the wide-field images.
A technique which computes the shear response only from the deep-field images would satisfy this
constraint. It would also allow us to easily run analyses on the wide-field shape catalog in the same
way as has been done in the past \citep[see, e.g.,][]{desy3shear} -- we'd compute our shear statistics
with the wide-field shape catalog, compute the response by applying the same selections to the
deep-field catalogs and applying Equation~\ref{eq:avgR}, and then apply the response to the data.
Second, we will still need the sheared noise correction term when computing the response from the
deep-fields, so we expect the final technique will double the pixel variance of the deep-field images.
Third, as we just argued, we'll need to match the PSF and noise distributions in the deep- and
wide-field images in order to correctly account for common sources of bias in the shear estimates.
A simple approach to match the noise distributions would be to apply a copy of the wide-field noise
to the deep-field image and vice versa. Finally, we'll need to
ensure that the various manipulations of the images with the PSF are numerically stable. Given that
we expect to need to match the PSFs between the deep- and wide-field images, a natural choice for
a reconvolution PSF for numerical stability would be the largest of the reconvolution PSFs of either
the deep- or wide-field images. \dmcal uses variants of these ideas as we show next.

The set of deep- and wide-field images, $\hat I_d$ and $\hat I_w$, that satisfies these constraints for an
artificial shear $\epsilon$ is
\begin{eqnarray}
  \hat I_{w} & = & I_{w}/P_{w} \star P_{r} + C_{w}\label{eqn:wide_im}\\
  \hat I_{d} & = & M[I_{d}, N_{d}', P_{d}, P_{r}, \epsilon] + C_{d}\label{eqn:deep_im}
\end{eqnarray}
where
\begin{equation}\label{eqn:cw}
  C_{w} = M[N^{''}_d, N^{'}_d, P_d, P_{r}, \epsilon=0]
\end{equation}
\begin{equation}\label{eqn:cd}
  C_{d} = N'_{w}/P_{w} \star P_{r}\ .
\end{equation}
Here $I_{w,d}$ are the input wide/deep-field images, $P_{w,d}$ are wide/deep-field PSFs,
$N^{',''}_{w,d}$ are wide/deep-field noise realizations, and $P_r$ is the largest reconvolution PSF
between the wide- and deep-fields. Expanding these equations out we find that the noise in
both images $\hat I_{w,d}$ can be computed as
\begin{equation*}
\hat N_{\alpha} = N_{w}/P_{w}\star P_{r} + Y[N_{d}/P_{d}, \alpha] \star P_{r} + Y[N'_d/P_d, -\alpha] \star P_{r}
\end{equation*}
with $\alpha\in\{0,+\epsilon,-\epsilon\}$. As is clear, we've avoided adding an extra
wide-field noise image at the cost of adding two extra deep-field noise realizations,
our original goal above. We've also matched the noise distributions in the images in the
same way as standard \mcal. Specifically, the wide-field (i.e., \noshear) image has noise
$\hat N_{\alpha=0}$, and the \onep and \onem (or \twop and \twom) images has noise
$\hat N_{\alpha=\pm\epsilon}$. Finally, we've also managed to match the PSFs in the
final images to $P_{r}$ and ensured numerical stability by choosing $P_r$ to be big
enough to suppress deconvolution noise effects from either $P_w$ or $P_d$.

The \dmcal shear estimate is computed via
\begin{equation} \label{eq:dmcalest}
  \hat{\gamma} \equiv \langle \mathbf{R} \rangle^{-1}_{\rm deep} \langle \mathbf{e} \rangle_{\rm wide}\ .
\end{equation}
Specifically, one computes the \dmcal response $R$ from the deep-field image set $\hat I_d$ and the
uncalibrated shear signal from the wide-field image set $\hat I_w$. In each set of images, one makes
identical selection cuts and shape measurements on the objects, and then computes the average shape over
the catalog of selected objects. Further details on how to generalize this procedure to a realistic survey
are given in Section~\ref{sec:statmatch}.

Figure~\ref{fig:dmcal} shows the various parts of the \dmcal images for \noshear, \onep,
and \onem cases. The left columm shows the pair of images with deep-field noise levels
and the middle column shows parts with wide-field noise levels. The pair of images comes from the two images
added together in the \mcal process, the original image and the noise image, in order to correct for sheared
background noise. The images in left and middle columns are added together to produce the final \dmcal images
$\hat I_{w,d}$ in the right column. We have exaggerated the artificial shear $\epsilon$ so that one can see the sheared
background noise and the oppositely sheared background noise correction for the deep-field
\mcal images in the last two rows of the left column.

In the top row for the \noshear image, the left column shows a pair of images that,
when added together, form $C_{w}$. These images add in noise levels and correlations
from the deep-field image put through \mcal. Thus effects like noise bias on the shears
or selections on signal-to-noise will be the same in both the wide- and deep-field images.
the middle column shows $I_{w}/P_{w} \star P_{r}$ which is the wide-field image that carries
the uncalibrated shear signal but PSF-matched to the final reconvolution PSF. Remember that
this reconvolution PSF $P_r$ is the same for both the wide- and deep-field images. Thus
this term serves to bring the wide-field PSF to the final PSF of the deep-field image
$\hat I_{d}$. The \mcal response of a galaxy to shear depends critically on the PSF,
in terms of both its value (e.g., larger galaxies relative to the PSF respond more
to shear), and through more subtle effects like selection, blending, and detection. By
matching the PSFs of the images, we ensure that these effects are the same in both the
wide- and deep-fields.

The middle and bottom rows show the \onep and \onem cases respectively. The left column
with the pair of images is the term $M[I_{d}, N_{d}', P_{d}, P_{r}, \epsilon]$. This term is
standard \mcal applied to the deep-field images. More explicitly, this term has the artificially sheared
image and the correction for the sheared background noise. These two items are needed for proper
shear response computations. The middle column shows $C_{d}$, which serves to bring the noise in
\dmcal response images to the wide-field noise level so that selections and noise bias levels
match between the wide- and deep-field images.

\begin{figure*}
    \centering
    \includegraphics[width=\linewidth]{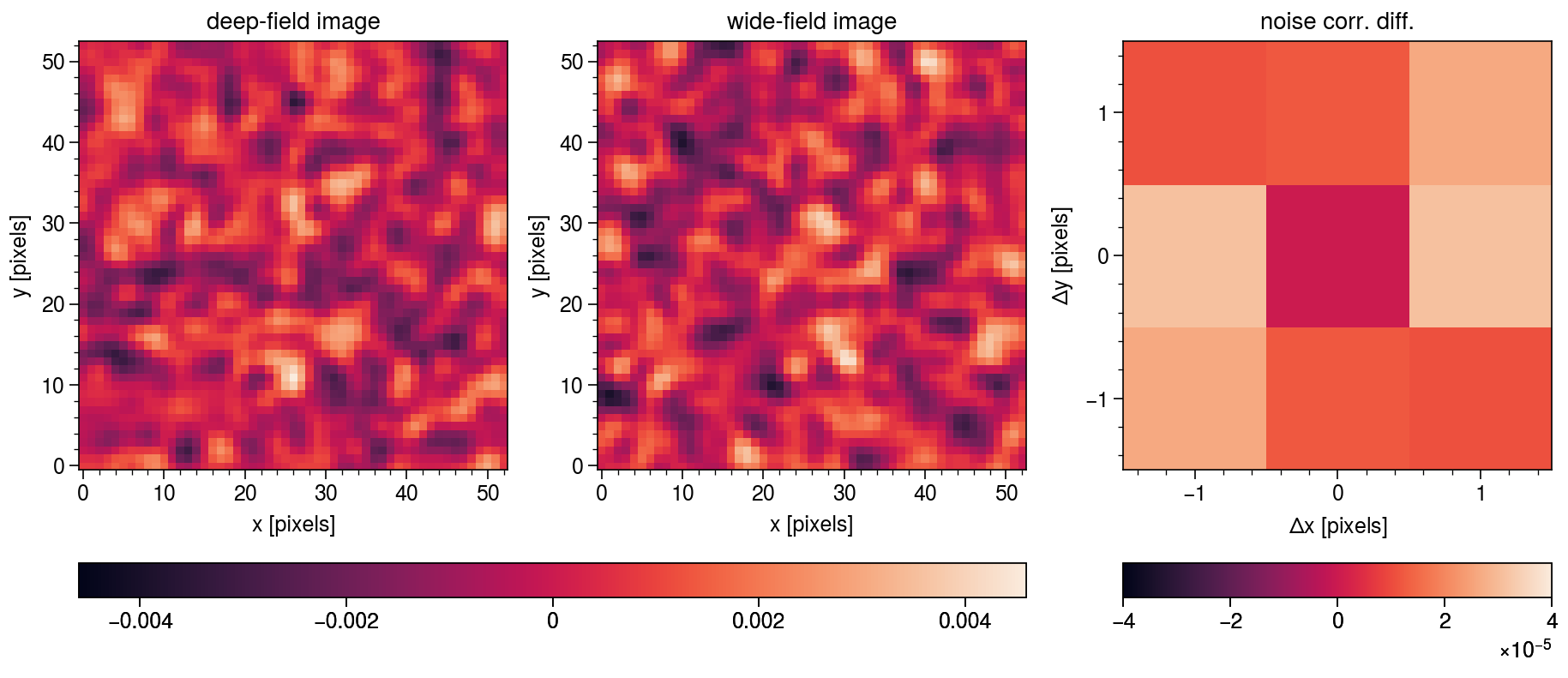}
    \caption{
      \textit{Left \& middle panel}: Example pure noise images from the deep- and wide-field image
      sets after \dmcal. Visually, the two noise images exhibit the same noise correlation structure.
      \textit{Right panel}: The difference in the average pixel-pixel correlation matrix of $10^6$ deep- and
      wide-field images $\hat I_{w,d}$. The difference is of order $10^{-5}$, indicating good numerical
      performance of our codes.
    } \label{fig:pixel_correlation}
\end{figure*}

In Figure~\ref{fig:pixel_correlation}, we compare the correlated noise fields for the
wide- and deep-field images, $\hat I_{w,d}$. The left and middle panels show example
realizations of the noise fields. Visually, the correlation structure is similar. To
test their agreement numerically, we compute the pixel-pixel correlation matrix for
adjacent pixels in both images, averaging over $10^6$ realizations to reduce the noise.
We show the difference between the wide- and deep-field pixel-pixel correlation matrices
in the right panel. They agree to ${\cal O}(10^{-5})$, indicating that our numerical
implementation of this technique is working well. We find similar fractional agreement
in the overall pixel variance.

\subsection{Processing Realistic Sets of Wide- and Deep-field Images}\label{sec:statmatch}

In a real survey, we will have a small set of deep-field images and a larger set of wide-field
images. More specifically, we won't have a deep-field observation of every wide-field object.
Thus we cannot naively apply the algorithm from the previous section to each object in the wide-field
catalog in order to compute the uncalibrated shear signal and its response. Importantly, \mcal-like
estimators only require that shape measurements in the numerator for $\langle \mathbf{e} \rangle$
and the denominator for $\langle \mathbf{R} \rangle$ match statistically. This matching is in terms
of not only the galaxy populations, but also the key image properties like PSF properties and noise levels.
\mcal ensures this statistical matching by using exactly the same set of objects/observations
for computing $\langle{\bf R}\rangle$ and $\langle{\bf e}\rangle$.
For \dmcal, we cannot rely on the wide- and deep-field image sets being statistically
identical \textit{a priori}. Instead, we will next devise a scheme to bring the two image sets into statistical
agreement before we attempt to measure the shear signal and its response. The most obvious way in which this
scheme might fail is sample variance. We estimate those effects in Section~\ref{sec:sv} below, demonstrating
that a Rubin LSST-like survey should not be limited by sample variance.

Our scheme works as follows.
\begin{enumerate}
\item For every wide-field image
\begin{enumerate}
  \item Draw a deep-field image at random with replacement.
  \item Use Equations~\ref{eqn:wide_im} and \ref{eqn:cw} to compute $\hat I_{w}$
    for the \noshear and use Equations~\ref{eqn:deep_im} and \ref{eqn:cd} to
    compute $\hat I_{d}$ for the $\pm\epsilon_{1,2}$ (i.e., \onep, \onem, etc.) cases.
  \item Perform shape measurements on each of
    the five images (\noshear, \onep, \onem, \twop, and \twom) from the last step.
  \item Accumulate a catalog of all shape measurements separately for each of the
    five images.
\end{enumerate}
\item Apply the same set of selection cuts separately to each of the five catalogs from
  the last step.
\item Use the mean shapes of each of the five catalogs separately after selection in
  order to compute the response corrected shear via Equation~\ref{eq:dmcalest}.
\end{enumerate}
This procedure is a brute-force Monte Carlo integral over the deep-field image
properties for each wide-field image. More specifically, four of the five catalogs above
will be of deep-field objects with properties computed from $\hat I_d$. The fifth catalog
will be of wide-field objects with properties computed from $\hat I_w$. We apply the same
selection cuts to all of the catalogs. The PSF and noise properties of these catalogs
are statistically identical, so that these selection cuts act in the same way in both
the deep- and wide-field datasets. These selection cuts can include any selection one
might want to make, including tomographic binning. Thus this procedure defines how to select
statistically equivalent sets of objects in the deep- and wide-fields in order to do a
realistic analysis. Due to the PSF and noise matching, the overall effects
of blending and source detection are the same as in standard \mcal. Namely, there will be
catastrophic shear biases for LSST-like surveys \citep{SheldonMdet2020}. We defer
exploring these blending and detection effects to future work and focus in this work
only on isolated objects, verifying that \dmcal works correctly in the same regime
as \mcal does.

In practice, there may be several avenues for improvement on the naive brute-force
approach taken above. First, as we show below, \dmcal works best when the PSF
of the deep-field images is the same size or smaller than the PSF of the wide-field
images. Thus it may be advantageous to subselect deep-field images by their PSF size
before pairing them to the wide-field images. A danger with this alternative procedure
is any potential inadvertent additional selections applied to the objects that cannot be
corrected by \dmcal (because they happen before \dmcal is applied). Another danger is
increasing the sample variance since now effectively only a subset of the deep-field
image set is used for each wide-field image. In practice, when coadded over hundreds or
epochs, the wide- and deep-field images in surveys like LSST will likely have very
similar, relatively narrow PSF distributions. Given that \dmcal is still
effective even if the deep-field PSF is marginally bigger than the wide-field PSF, this
alternative matching procedure may only supply marginal gains.

Second, the procedure defined above produces one deep-field image for every wide-field
one. In practice, since the deep-field survey is smaller in area than the wide-field
survey, this will mean much of the deep-field data is duplicated (but with different
wide-field noise levels and reconvolution PSFs applied). It may be more efficient to
separately produce much smaller deep-field image sets by using an algorithm where for
each deep-field image, one selects at random a relatively small number of wide-field
images to compute $C_{w}$. One must be careful to avoid extra noise from incomplete
wide-field image sampling in this procedure, but this is easily tested in simulations of
shear recovery. This change would produce smaller object catalogs for the
deep-field measurements, making downstream manipulations of the catalogs easier.

Third and finally, we have made no attempt to reduce sample variance in the deep-field
response computation by matching the properties of the wide- and deep-field objects.
This matching should be possible within the \dmcal formalism. The idea is to define
subsets of the wide-field objects through selection cuts on the
\dmcal-computed properties. These subsets should group similar objects by size,
signal-to-noise, etc. Then in the deep-field samples, one would apply the same
selections to the \dmcal-computed properties. For each subset, one can compute
a separate response. Then the total wide-field sample response can be computed by an
appropriately weighted average of the deep-field response values. This reweighting
procedure should cause no additional biases in the \dmcal shear estimator
while reducing the overall sample variance in the response estimate.

One appealing aspect of \dmcal is that it naturally generalizes to
\mdet. In \dmdet, we process larger images through the \mcal procedure, not
just postage stamps, and perform both object
detection and object shape measurement separately for all \mcal sheared
and unsheared images. For \dmdet, we simply apply the
same set of steps above except that we use larger images and run object detection before
shape measurement. The measured object shapes are accumulated into catalogs in the same way.
The response after object selections is computed in the same way as well. Further, all
of the possible improvements discussed above also generalize to \dmdet. This algorithm will
fully handle the effects of blending and object detection. We leave explicit tests of the improvements discussed above and \dmdet to
future work.

Below we test a typical setup for standard \mcal where each image is a
postage stamp around a single wide- or deep-field galaxy, ignoring both detection and blending.
Also, in order to separate out the effects of sample variance from our main results,
we generate one deep-field image with a new galaxy chosen at random for every wide-field
image, effectively working in the limit that the wide- and deep-field data cover
equal but independent areas of the sky. Finally, we work in the limit of perfectly known
PSF and noise properties for the deep- and wide-field images, as has been the standard in
past works \citep[see, e.g.,][]{SheldonMcal2017,SheldonMdet2020,SheldonMdet2022,li2018fpfs,li2022fpfsblend}.
Future work should carefully consider the effects of PSF and noise misestimation errors for a
realsitic survey analysis pipeline.

\section{Results}\label{sec:results}

In this section we present our main results. These include tests of shear recovery
(Section~\ref{sec:mc}, Table~\ref{tab:shearmeas}) and estimates for how effective the
technique is at reducing noise as a function of the relative exposure times and PSF sizes of the wide- and
deep-field images (Section~\ref{sec:sizedepth}, Figure~\ref{fig:s2n}). We discuss the
relative importance of the various correction images used by \dmcal in
Section~\ref{sec:terms}. We conclude by estimating the effects of sample variance on our
estimator using two complementary simulated models of galaxy populations
(Section~\ref{sec:sv}, Figure~\ref{fig:sample_variance}).

\begin{table*}
  \centering
  \begin{threeparttable}
  \caption{Shear Calibration Results for \DMcal}
  \label{tab:shearmeas}

  \begin{tabular}{cccccc}
    \hline
    \noalign{\vskip 1mm}
    case & noise\tnote{a} & PSF\tnote{b} & galaxy\tnote{c} & m [$10^{-3}$, $3\sigma$ error] & c [$10^{-4}$, $3\sigma$ error]\\
    \noalign{\vskip 1mm}
    \hline
    \noalign{\vskip 1mm}
    1 & fixed    & fixed Gaussian    & exponential & $0.37\pm0.17$   & $\phantom{-}0.09\pm1.38$  \\
    2 & variable & fixed Gaussian    & exponential & $0.36\pm0.23$   & $-0.62\pm1.77$ \\
    3 & variable & variable Gaussian & exponential & $0.30\pm0.20$   & $\phantom{-}0.15\pm0.90$  \\
    4 & variable & variable Gaussian & \descwl     & $0.53\pm0.39$   & $-0.22\pm0.64$  \\
    5 & fixed    & fixed Moffat      & exponential & $0.43\pm0.20$   & $\phantom{-}0.09\pm1.28$  \\
    6 & variable & fixed Moffat      & exponential & $0.14\pm0.21$   & $\phantom{-}0.18\pm1.36$ \\
    7 & variable & variable Moffat   & exponential & $0.38\pm0.22$   & $\phantom{-}0.68\pm0.94$  \\
    8 & variable & variable Moffat   & \descwl     & $0.53\pm0.37$   & $\phantom{-}0.00\pm0.64$ \\
    \noalign{\vskip 1mm}
    \hline
  \end{tabular}

  \begin{tablenotes}
  \item [a] All simulations use a deep-field exposure time that is on average $10\times$ more
    than the wide-field. \textit{fixed} noise uses a fixed noise level for the
    wide- and deep-field images, such that the deep-field noise level is a factor of
    $1/\sqrt{10}$ lower than the wide-field noise level. \textit{variable} noise
    applies an additional uniform random factor between 0.9 and 1.1 to the noise levels
    of the both the wide- and deep-field images.
  \item [b] The \textit{fixed} PSF model assumes the wide- and deep-field PSFs are
    fixed Gaussian or Moffat profiles with FWHMs of 0.9'' and 0.7'' respectively. The \textit{variable} PSF
    model assumes the wide- and deep-field PSFs are both Gaussian or Moffat profiles with FWHMs drawn from
    uniform distributions between 0.8'' to 1.0'' and 0.6'' to 0.8'' respectively.
  \item [c] The exponential galaxy model is an object with an exponential profile with
    half-light radius of 0.5''. The \descwl model draws bulge+disk objects to match an
    LSST-like survey using the \descwl package
    \citep{WeakLensingDeblendingPaper,WeakLensingDeblendingSoftware}.
  \end{tablenotes}
  \end{threeparttable}
\end{table*}

\subsection{Shear Recovery}\label{sec:mc}

In order to test shear recvovery, we simulate objects in postage stamps using the
\galsim package. The wide- and deep-field objects parameters are drawn from the same distributions
and we work in the limit of an unlimited amount of deep-field data in order to reduce
the noise and test the intrinsic accuracy of our methods. We further use the techniques
from \citet{pujol2019} to cancel the pixel-noise in the simulations by generating pairs
of deep- and wide-field objects with the same pixel-noise but opposite true shears. See
\citet{SheldonMdet2020} for an explicit discussion of how these techniques are applied
to \mcal and \mdet. All shape measurements are done with a post-PSF 1.2'' Gaussian
weighted moment. We keep only objects with signal-to-noise greater than ten and
$T/T_{\rm psf} > 1.2$, where $T=\sigma_{x}^2 + \sigma_{y}^2$ from the moment measurement
of the galaxy or PSF. We report our results in terms of the standard additive and
multiplicative bias parametrization \citep[see, e.g.,][]{heymans2006}
\begin{equation*}
g_{\rm meas} = (1+m)g_{\rm true} + c
\end{equation*}
where $m$ is the multiplicative bias and $c$ is the additive bias. All reported errors
are $3\sigma$.

We present the results of our shear recovery tests in Table~\ref{tab:shearmeas} for
various cases. We have introduced additional complications into the process in a staged
fashion in order to ensure we can identify the source of biases, even if
they may cancel to some degree in our final results. We start with
case 1, a simple setup using exponential galaxies with half light radius of 0.5'', a
Gaussian PSF with FWHM 0.9'' for the wide- field images, and a Gaussian PSF with FWHM
0.7'' for the deep-field images. The noise level is set to generate objects with a
typical signal-to-noise of $\approx19$ for the wide-field images. The deep-field images
are set to have 10$\times$ less noise variance than the wide-field images ($\sqrt{10}$
smaller standard deviation).
Both the wide- and
deep-field images are $53\times53$ postage stamps with a 0.2" pixel scale.
We find that our
techniques can recover the true shear correctly ($m=0.37\pm0.17$ [$10^{-3}$, $3\sigma$
error], $c=0.09\pm1.38$ [$10^{-4}$, $3\sigma$ error]) up to second-order shear effects
which are expected to be a few parts in ten thousand
\citep{SheldonMcal2017,SheldonMdet2020}.

For case 2, we add variations in the noise levels within the wide- and deep-field
samples. For this test case, we use the same setup as the previous test case, but scale
the wide- and deep-field noise levels by a random uniform draw between 0.9 and 1.1. This
additional test case serves to ensure that we are correctly matching the image noise
properties with our technique. If we did not match the noise properties precisely, we'd
expect to see biases in our results. We again find results consistent with second- order
shear effects, getting $m=0.36\pm0.23$ [$10^{-3}$, $3\sigma$ error] and $c=-0.63\pm1.77$
[$10^{-4}$, $3\sigma$ error].

Case 3 introduces PSF size and shape variations into the previous noise variations test
case. For this we draw the wide- field PSF FWHM from a uniform distribution between
0.8'' and 1.0''. The deep-field PSF FWHM is drawn from a uniform distribution between
0.6'' and 0.8''. We also apply a small random shear to the PSFs, drawn uniformly
between -0.02 and 0.02 for each component. This test case ensures that we are appropriately PSF matching the
wide- and deep-field samples. If the PSF sizes or shapes were mismatched, we'd expect to
see increased multiplicative and additive biases. Again we find results consistent with
second-order shear effects, getting $m=0.30\pm0.20$ [$10^{-3}$, $3\sigma$ error] and
$c=0.14\pm0.90$ [$10^{-4}$, $3\sigma$ error].

For case 4, we introduce a population of realistic galaxies. This test case
serves to ensure that the statistical matching of wide- and deep-field objects and noise
properties via the procedure in Section~\ref{sec:statmatch} is working correctly. For
this case we use the same setup as the last simulation but replace our exponential
galaxies with randomly drawn bulge+disk objects from the \descwl package
\citep{WeakLensingDeblendingPaper,WeakLensingDeblendingSoftware} . We match the
wide-field noise level to a 10-year LSST survey, applying a uniform variation factor
between 0.9 and 1.1 as before. The deep-field exposure time is set to be 10$\times$ more than
the wide-field and to the image noise is scaled by a factor of $1/\sqrt{10}$.
The deep-field noise level is also varied uniformly by a factor between 0.9 and 1.1.
Unlike the previous cases, we simulate the deep-and wide-field images using
$73\times73$ postage stamps with the same 0.2" pixel scale. The increased image dimension allows
for the larger objects in the \descwl catalog. We again find results consistent
with second-order shear effects, with $m=0.57\pm0.39$ [$10^{-3}$, $3\sigma$ error] and
$c=-0.13\pm0.64$ [$10^{-4}$, $3\sigma$ error]. This test neglects the effects of sample
variance since we have drawn one deep-field galaxy for every wide-field galaxy. We will
explore the effects of sample variance in Section~\ref{sec:sv} below.

Finally, we have repeated the tests above with Moffat profiles from the \galsim package. We use
a shape parameter $\beta=2.5$ and the same range of PSF FWHM values for each case. The Moffat
profile has larger tails than a Gaussian and so serves to test the effectiveness of our reconvolution
kernel algorithms. The results of these tests are listed as cases 5-8 in Table~\ref{tab:shearmeas}.
We again find no biases beyond second-order shear effects.

\subsection{Effectiveness as a function of Relative PSF Size and Exposure Time}\label{sec:sizedepth}

Now that we have established the accuracy of our technique, we turn our
attention to the expected gains in realistic survey scenarios where deep field
can have larger PSFs than the wide field, and when the relative exposure times of the
wide- and deep-fields vary.

We first show results for the PSF variation.  For this setup, we simulate
exponential galaxies with half light radii of 0.7'' with the wide-field PSF size varying
between 0.7'' and 1.1''. We take the deep-field to have $10\times$ the exposure time of the
wide-field and vary the ratio of the deep-field to wide-field PSF size from 0.5 to 1.5.
The rest of the simulation parameters follow case 1 above in Section~\ref{sec:mc}.
For each configuration, we measure the ideal signal-to-noise of
our galaxy, the signal-to-noise with standard \mcal, and the signal-to-noise with
\dmcal. We define the ideal signal-to-noise as that from the original wide-field image
of the object with no \mcal or \dmcal operations applied.

\begin{figure*}
    \centering
    \includegraphics[width=\textwidth]{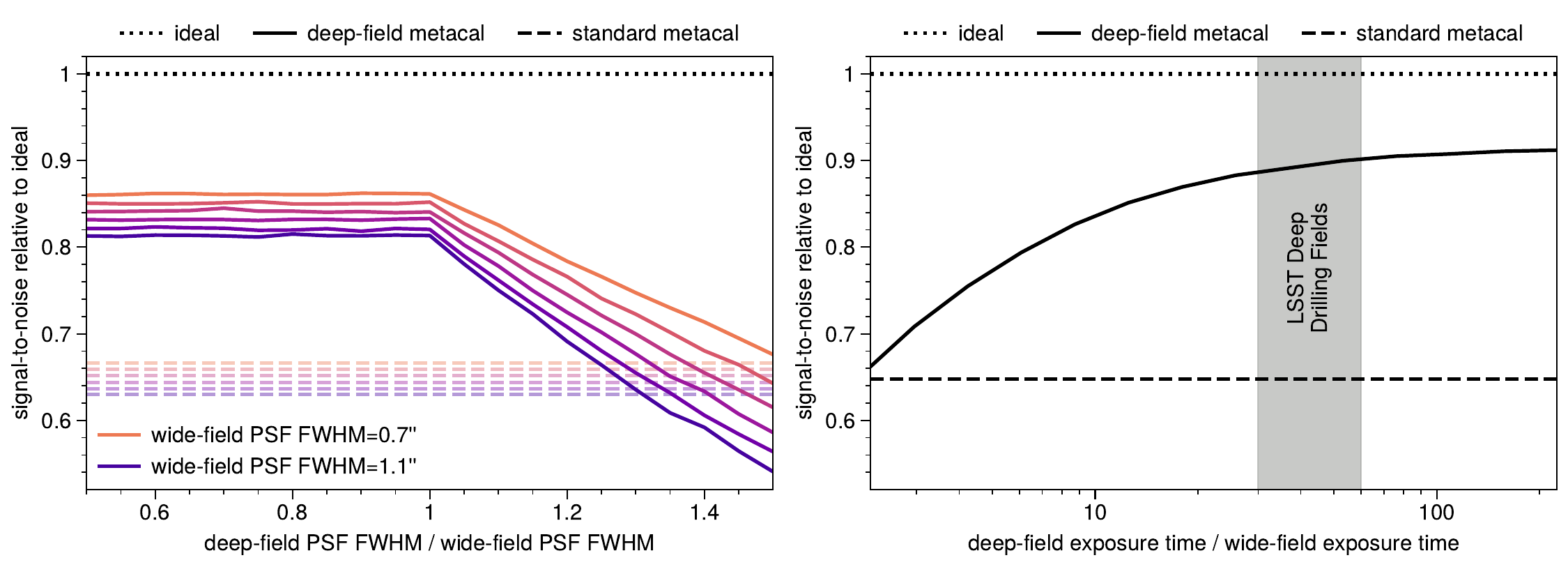}
    \caption{
      \textit{Left panel}: Signal-to-noise relative to ideal of sources processed deep-field and standard \mcal
      as a function of the ratio of deep-field PSF FWHM to the wide-field FWHM. We assume the deep-field has
      10$\times$ the exposure time of the wide-field for this plot. The different colored solid lines show various wide-field PSF FWHMs ranging from 0.7'' to
      1.1'' for \dmcal. The lower dashed lines of different colors show the result at the same wide-field PSF FWHM for
      standard \mcal. The dotted line is the ideal result of one.
      \textit{Right panel}: Signal-to-noise relative to ideal of sources processed with \dmcal and standard \mcal as a function of the relative exposure time of the deep- versus wide-field observations. The FWHMs of the PSFs are fixed at 0.9''
      and 0.7'' for the wide and deep images. The solid line shows \dmcal and the lower dashed line shows standard \mcal.
      The dotted line is the ideal result of one.
      The solid gray band shows the estimated relative exposure time of the LSST Deep Drilling fields compared to the main LSST survey.
      Notice that in both the left and right panels, the maximum signal-to-noise for \dmcal is limited to roughly 90\% of
      the ideal case. This effect is partly due to the increased size of the \mcal reconvolution PSF and our use of a fixed aperture moment
      for flux and shape measurement.}
    \label{fig:s2n}
\end{figure*}

The results of this test are shown in the left panel of Figure~\ref{fig:s2n}, where we
plot the signal-to-noise in units of the ideal signal-to-noise. The dashed horizontal
lines at the bottom of the figure show the results of standard \mcal. The colored
shading in the lines shows results for PSF FWHMs varying between 0.7'' and 1.1''. As
expected for the fixed, post-PSF aperture moments employed in this work, larger PSF
sizes reduce the overall signal-to-noise. The solid lines show the results of deep-field
\mcal. When the deep-field PSF is of the same size or smaller than the wide-field PSF,
we see substantial gains in the signal-to-noise. This gain matches our expectations from
the math above. The signal-to-noise for \dmcal is about 30\% higher than for \mcal,
$0.85/0.65\approx1.3$, when comparing the dashed and solid lines on the left side of
the plot. For this ratio of signal-to-noise measures, we expect
$\sqrt{2\sigma^2_{w}/(\sigma^2_{w} + 2\sigma^2_{d})}=\sqrt{2/(1 + 2/10)}\approx1.30$ since \dmcal
trades one copy of the wide-field noise $\sigma^{2}_{w}$ for two copies of the deep-field
noise $\sigma^{2}_{d}$. Notice however that even in the case where the
deep-field PSF is moderately larger than the wide- field PSF, \dmcal is still
lower noise than standard \mcal. The average increase in signal-to-noise per object
will result in weak lensing samples from \dmcal that extend to higher redshift
than those from standard \mcal. We leave estimating the amplitude and significance of
this effect to future work.

The final salient feature of the plot is that even if \dmcal is working its
best, we still find the signal-to-noise to be about $10\%$ lower than the ideal case,
shown as the dotted line. This feature is partly due to the increased size of the reconvolution
PSF used in the \dmcal image manipulations. We have used a relatively
conservative setting of the reconvolution PSF size in this work. Better techniques to
set the reconvolution PSF size, like those employed in the Dark Energy Survey \mcal
shear measurements \citep{desy3shear} would result in smaller losses due to this effect.

Now we turn to the effects of the relative exposure time of the wide- and deep-field images. For
this test, we fix the sizes of the wide- and deep- field PSFs to 0.9'' and 0.7''
respectively and vary the exposure time of the deep-field image from 2$\times$ to
$\sim200\times$ more than the wide-field image. The rest of the simulation
parameters follow case 1 from Section~\ref{sec:mc}. We again measure the
signal-to-noise in units of the ideal signal-to-noise for our test galaxy. The results
of this computation are shown in the right panel of Figure~\ref{fig:s2n}. The dashed
line in this panel shows the results of standard \mcal. The solid line shows the results
for \dmcal. The dotted line shows the ideal result of one. We find, as
expected, that as the exposure time of the deep-field image increases, our
technique becomes more effective. Again we see an overall loss of $\sim10\%$ in
signal-to-noise due to the reconvolution PSF. In the gray band, we show the estimated
exposure time of the LSST Deep Drilling Fields relative to the main LSST survey \citep[DDF,][]{lsst-ddf-depth}. While
the actual achieved exposure time of these fields is not known yet, for a broad range of
plausible exposure times, \dmcal will be quite effective at reducing noise.

While the results in this section demonstrate the benefits of \dmcal for
specific sources, the overall increase in the precision of a weak lensing analysis with
\dmcal depends on the exact decreases in source shape noise and increases in
the effective number density due to the reduction in pixel noise. Further, for a survey
like the Rubin LSST, the exact area and exposure time the DDFs are not yet known. Similarly, we
do not yet know the relative PSF properties of the DDFs versus the main LSST survey. In
order to proceed, we make the following relatively conservative approximation. We assume
that the subset of the DDF images with the smallest PSF sizes are used to reach an exposure time
of only $10\times$ that of the main LSST survey. We then simulate objects as in case 4
above using the \descwl package but forcing the DDFs and the main LSST survey
to have the same PSF FWHM of 0.8''. We also fix the relative exposure time between the two and
neglect the PSF shape variations from case 4 to reduce noise in our estimate. For each
object, we measure its shear using standard \mcal and \dmcal. We then compute
the error on the mean shear which is dictated by the same combination shape noise and
effective number density that enters the cosmic shear covariance matrix (i.e.
$\sqrt{\sigma_e^2/n}$ where $\sigma_e^2$ is the shape noise and $n$ is the number density
of sources). We find that \dmcal achieves an $\approx15\%$ smaller
error on the mean shear relative to standard \mcal.  Note that
\citet{SheldonMcal2017} found that the increased pixel noise in standard
\mcal resulted in an $\approx 20\%$ overall degradation in the precision of
statistical weak lensing measurements relative to an ideal case with no increased pixel
noise.  Our results thus suggest that the degradation
is reduced to a few percent with \dmcal.  As our simulations do
not contain perfect realism, we conservatively estimate the degradation for deep-field
\mcal is $\lesssim5\%$.  In more realistic scenarios, we may be able to use more of the DDF
data resulting in better performance. We defer more detailed estimates to future work.

\subsection{Relative Importance of Different Correction Terms}\label{sec:terms}

In this section, we explore the relative importance of the two correction images,
Equations~\ref{eqn:cw} and \ref{eqn:cd}. We consider a setup like case 1 above, but we
remove one or both of these terms before computing the mean shear. For this case we also
match the PSFs of the wide- and deep-fields, keeping them both at 0.8''. We also use
sources with signal-to-noise of $\approx12$ to accentuate the effects of differing noise
on the shear calibration. We find that without the wide-field image correction term, the
shear recovery is biased with $m=37.2\pm8.1$ [$10^{-3}$, $3\sigma$ error]. Similarly
without the deep-field image correction term, we again find that the shear recovery is biased
with $m=-43.5\pm2.2$ [$10^{-3}$, $3\sigma$ error]. Finally without either term, we find
$m=-6.1\pm2.2$ [$10^{-3}$, $3\sigma$ error], indicating some degree of cancellation in
the various effects at play. The differing signs in the biases reflect changes in the
response to shear of the numerator of the \dmcal estimator versus the
denominator. When removing the extra deep-field noise applied to the wide-field image
(i.e., Equation~\ref{eqn:cw}), the wide-field shape measurements respond more to shear
than the deep-field shape measurements, generating positive biases. Similarly, without
the extra wide-field noise applied to the deep-field image (i.e.,
Equation~\ref{eqn:cd}), the deep-field shape measurements respond more to shear than the
wide-field ones, generating negative biases. The two effects largely cancel, but leave
biases that exceed LSST requirements by a factor of several. These results indicate that
in order to make high-precision shear measurements, the noise distributions in the
images must be treated carefully.

\subsection{Sample Variance}\label{sec:sv}

We now address the issue of sample variance in the deep-field dataset. The LSST DDF will
cover $\approx35$ deg$^2$ of area \citep{lsst-ddf-design,ivezic2019lsst}. In deep-field
\mcal, we compute the survey response only from the deep-field area. Thus sample
variance in the galaxy populations for this area will imprint itself in the \mcal
response, effectively increasing our final error on the multiplicative bias. This effect
is most easily characterized by the fractional scatter in the deep-field response
$\langle R_{d}\rangle$. This fractional scatter is directly comparable to our
requirements on the knowledge of the multiplicative bias $m$. For an LSST-like survey,
the requirement for the final 10-year data is $m \lesssim 0.1-0.2\%$
\citep{huterer2006,descsrd}.

In order to estimate the magnitude of this effect, we have used two different structure
formation models to compute the fractional scatter in the deep-field response as a
function of the deep-field area. The first is the \buzzard simulation set
\citep{derose2019buzzard,derose2021Buzzard} which uses the \textsc{ADDGALS} \citep{addgals} algorithm to
populate low-resolution N-body simulations with galaxies as a function of luminosity
and color. This method is able to provide $\approx5,000$ deg$^2$ of mock galaxy
populations. These catalogs match a variety of structure formation statistics at low
redshift, but their ability to make high-redshift predictions is not fully tested. The
\buzzard catalogs do not provide object shapes or sizes that correlate with large-scale
structure. To add in this effect, we use a simple matching technique, the basis of which
is described in \citet{galsampler}. We apply a linear transformation to the \buzzard
catalog to match the median and standard deviation of the $r-i$ and $i-z$ colors  of
objects in the \descwl catalog. To generate a specific deep-field data realization, we
first select a contiguous region of some area from the \buzzard catalog. We then find
the closest 100 galaxies from the \descwl catalog to each galaxy in the \buzzard catalog
in $r-i$, $i-z$ color-color space. We then select one of the 100 closest galaxies at
random, and use it in simulating our deep-field sample for this area. This process
imprints sample variance in the galaxy populations from \buzzard into the \descwl
catalog. The second model is the public \cosmodctwo \citep{cosmodc2} catalog. It
simulates a full set of galaxy properties in a 440 deg$^2$ area, including object shapes
and sizes that correlate with large-scale structure. This catalog has been used
extensively by the LSST Dark Energy Science Collaboration for simulating LSST-like
observations \citep{dc2,dc2note}.

\begin{figure}
    \centering
    \includegraphics[width=0.45\textwidth]{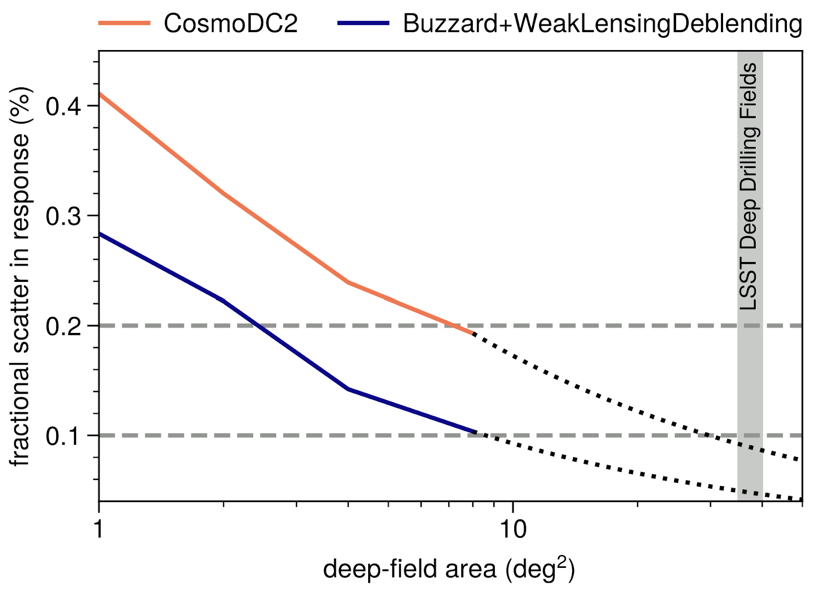}
    \caption{
      Sample variance scatter in the deep-field response as a function of the
      deep-field area. The different colored lines show the estimated sample variance
      from two different mock galaxy catalogs, \cosmodctwo and \buzzard+\descwl. The
      dashed lines represent the expected range of requirements on the shear
      calibration. The solid gray band shows the estimate area of LSST Deep Drilling
      Fields. The dotted lines show extrapolations to the LSST Deep Drilling Fields
      area assuming the sample variance scales as $1/\sqrt{\rm area}$. The LSST Deep
      Drilling fields will be large enough to result in sample variance scatter levels
      around 0.05\% to 0.1\%, meeting LSST requirements.
    }
    \label{fig:sample_variance}
    \vspace{1em}
\end{figure}

For both models, we simulate the deep-field samples assuming the deep-field data has
10$\times$ more expsoure time than the wide-field data. Due to the fact that the wide-field noise is
applied to the deep-fields, we cut the input simulation catalogs keeping only objects
brighter than $r$-band of $26$. For this specific test, we fix the PSF FWHM to 0.7'' for
both the wide- and deep-field images. All images are $73\times73$ pixel postage stamps
with a 0.2'' pixel scale. With this setup, we compute the fractional scatter
in the shear response for a variety of assumed deep-field areas, ranging from 1 deg$^2$
to 8 deg$^2$. The results of this computation are shown in
Figure~\ref{fig:sample_variance}. We find that the fractional scatter decreases
approximately as $\sim1/\sqrt{\rm area}$ as one might expect. The simulations give
somewhat different results, with \cosmodctwo consistently about 0.1\% higher than the
\buzzard+\descwl model. For areas approaching that of the LSST DDF, we estimate a
residual fractional scatter in the range of 0.05\% to 0.1\% assuming a DDF area of 35
deg$^2$ and the $1/\sqrt{\rm area}$ scaling. This level of sample variance will meet or
be less than LSST requirements.

We have examined the galaxy properties of the \cosmodctwo and \buzzard+\descwl samples
in order to understand why they give different results. We find that the \cosmodctwo
model has a broader variance in the bulge-to-disk ratio while the shape noise
distributions are largely similar. We also find that the \cosmodctwo catalog has a
slightly higher number density of objects, with $\approx65\ {\rm objects}/{\rm
arcmin}^2$ as opposed to the \buzzard+\descwl catalogs with only $\approx50\ {\rm
objects}/{\rm arcmin}^2$. This difference in number density does not account for the
differences in the response sample variance scatter. Whatever the exact nature of
the differences between the catalogs, our current results indicate that for a broad
range of assumptions about the LSST DDF samples, \dmcal will be an effective
technique.

\section{Summary}\label{sec:conc}

In this work, we have developed a new technique called \dmcal. It allows one
to combine images from deep- and wide-field data sets to measure weak lensing
gravitational shear using \mcal-type methods but at higher precision. In particular our main
conclusions are:
\begin{itemize}
  \item \Dmcal allows one to combine images from wide- and deep-field surveys
    to measure \mcal weak lensing shears at below part-per-thousand accuracy for isolated sources
    (Section~\ref{sec:mc}, Table~\ref{tab:shearmeas}) while reducing the pixel noise
    in \mcal images by $\approx30\%$ (Section~\ref{sec:sizedepth}).
  \item The exact decrease in pixel noise depends on the details of
    the wide- and deep-field exposure time and PSF distributions (Section~\ref{sec:sizedepth}).
  \item When applied to the Rubin LSST, we expect an at least $\approx15\%$ increase in
    the precision of weak lensing analyses due to the decreased pixel noise contributions
    to the shape noise and increased effective source density. We conservatively
    estimate that the corresponding degradation in the precision of statistical
    shear measurements relative to an ideal case of no extra pixel noise
    is reduced from 20\% for standard \mcal to $\lesssim5\%$ for \dmcal.
  \item The LSST DDFs will have enough area to suppress the effects of sample variance
    in the deep-fields to levels that will meet LSST requirements
    (Section~\ref{sec:sv}).
  \item \Dmcal will extend naturally to \mdet due to the fact that it
    explicitly matches the PSF of the deep- and wide-field images, ensuring detection
    and blending biases are properly calibrated (Section~\ref{sec:deepmcal}).
\end{itemize}

We have also discussed several avenues for improving the algorithms presented in this
work. These include optimizing how one pairs the wide- and deep-field images in order to reduce
signal-to-noise losses from PSF matching (Section~\ref{sec:statmatch}), reducing
computing costs through reducing duplication in the deep-field precessing
(Section~\ref{sec:statmatch}), and reducing sample variance in the deep-field response
computations through reweighting the deep-field catalogs within \mcal/\mdet to better
match the wide-field object distributions (Section~\ref{sec:statmatch}).
Finally, it is important that this work be extended to test \dmdet,
in order to work out technical issues related to
handling real-world effects like image coaddition, masking/artifacts, bright stars,
PSF mistestimation errors, noise misestimation errors, etc.
Detailed tests of \dmdet will be needed to ensure our techniques can reach
their full potential.

In future work, we will pursue the above improvements, perform explicit tests
of \dmdet for LSST-like surveys, and develop an implementation of
\dmdet for Rubin LSST data.

\section*{Acknowledgments}

MRB is supported by DOE grant DE-AC02-06CH11357.  ES is supported by DOE grant DE-AC02-98CH10886.
We thank Joe DeRose and Risa Wechsler for providing the \buzzard simulation used in this work.
We thank the \cosmodctwo team
for publicly releasing their mock galaxy catalogs. We thank Andrew Hearin and Eduardo
Rozo for extremely helpful comments that greatly improved the presentation of this work.
We gratefully acknowledge the computing resources provided on Bebop, a high-performance
computing cluster operated by the Laboratory Computing Resource Center at Argonne
National Laboratory, and the RHIC Atlas Computing Facility, operated by Brookhaven
National Laboratory. This work also used resources made available on the Phoenix
cluster, a joint data-intensive computing project between the High Energy Physics
Division and the Computing, Environment, and Life Sciences (CELS) Directorate at Argonne
National Laboratory. This work made extensive use of the Astrophysics Data Service (ADS)
and \texttt{arXiv} preprint repository. We thank the maintainers of the
\texttt{GCRCatalogs}, \citep{gcrcatalogs}, \texttt{numpy} \citep{numpy}, \texttt{scipy}
\citep{scipy}, \texttt{numba} \citep{numba}, \texttt{Matplotlib} \citep{matplotlib}, and
\texttt{conda-forge} \citep{condaforge} projects for their excellent open-source
software and software distribution systems.

\bibliographystyle{aasjournal}
\bibliography{references}


\end{document}